\documentclass[aps,pre,signlecolumn,showpacs,preprintnumbers,amsmath,amssymb]{revtex4}

\usepackage{mathrsfs}
\usepackage{graphicx}
\usepackage{dcolumn}
\usepackage{bm}

\begin{document}

\title{Heat conduction induced by non-Gaussian athermal fluctuations}

\author{Kiyoshi Kanazawa$^1$, Takahiro Sagawa$^{1,2,}$\footnote{Present Address: Department of Basic Science, The University of Tokyo, Komaba 3-8-1, Meguro-ku, Tokyo 153-8902, Japan}, and Hisao Hayakawa$^1$}

\affiliation{	$^1$Yukawa Institute for Theoretical Physics, Kyoto University, Kitashirakawa-oiwake cho, Sakyo-ku, Kyoto 606-8502, Japan\\
				$^2$The Hakubi Center for Advanced Research, Kyoto University, Yoshida-ushinomiya cho, Sakyo-ku, Kyoto 606-8302, Japan	
			}
\date{\today}

\begin{abstract}
		We study the properties of heat conduction induced by non-Gaussian noises from athermal environments. 
		We find that new terms should be added to the conventional Fourier law and the fluctuation theorem for the heat current, 
		where its average and fluctuation are determined not only by the noise intensities but also by the non-Gaussian nature of the noises. 
		Our results explicitly show the absence of the zeroth law of thermodynamics in athermal systems. 
\end{abstract}

\maketitle
\section{Introduction}
		Recent developments of experimental technologies have enabled us to investigate the detailed thermodynamic properties of small systems such as colloidal and biological systems~\cite{Bustamante}.
		If the environments of the systems are in thermal equilibrium, 
		stochastic thermodynamics with Gaussian noises has shown to be very powerful 
		to investigate universal relations in nonequilibrium statistical mechanics of small systems~\cite{Seifert5,Seifert4,Sekimoto1,Sekimoto2,Sekimoto3,Blickle}. 
		In these systems, for example, the average and the fluctuation of heat current are characterized by the Fourier law and the heat fluctuation theorem, respectively
		~\cite{Bonetto,Kubo,Chang,Li,Ciliberto,     Evans,Lebowitz,Crooks,Jarzynski,Kurchan,Seifert0,Zon,Nemoto,Noh,Harada,Dybiec,Andrieux}. 
		On the other hand, the effects of non-Gaussian noises from athermal environments have been reported in 
		electrical circuits~\cite{Blanter} and biomolecular systems~\cite{Ben}. 
		The conventional approaches in stochastic thermodynamics are not applicable to such systems, because the environments are not in thermal equilibrium. 
		An alternative approach to this problem has been the formulation in terms of non-Gaussian noises~\cite{Reimann,Luczka,Touchette1,Baule,Utsumi,Morgado1,Morgado2}. 
		However, a universal theory of nonequilibrium statistical mechanics in the presence of non-Gaussian noises has not been fully understood. 
		For example, how should the fundamental thermodynamic relations, such as the Fourier law and the heat fluctuation theorem, be modified with non-Gaussian noises? 
		
		In this paper, we answer this question with a stochastic model of heat conduction induced by non-Gaussian noises from athermal environments. 
		We derive generalizations of the Fourier law and the fluctuation theorem, by applying non-Gaussian stochastic energetics on the basis of a new stochastic integral introduced in Ref.~\cite{Kanazawa}. 
		The average heat current between the environments is determined not only by the difference in the noise intensities (i.e., the temperatures for the case of equilibrium environments), 
		but also by the difference in the non-Gaussianity of the noises. 
		In particular, even when the noise intensities of the environments are the same, the heat can be conducted purely by the effect of the non-Gaussianity. 
		We also derive a correction to the heat fluctuation theorem, which reveals the fundamental properties of the heat fluctuations in the presence of the non-Gaussian noises. 
		Moreover, we investigate the validity of the zeroth law of thermodynamics for athermal systems, 
		and find that the zeroth law is not universally valid but depends on the details of a contact device between two systems. 
		We numerically verify our statements, 
		which demonstrate that the direction of the average heat current depends on the characteristics of the heat conductor, 
		and that the properties of the heat fluctuation significantly deviates from those of the conventional fluctuation theorem. 
		Our result would serve as a theoretical foundation to study the energy transport and the irreversible phenomena in athermal systems with non-Gaussian noises. 
		
		This paper is organized as follow. 
		In Sec. II, we formulate the model of a Brownian motor with non-Gaussian noises, and define the heat current.  
		In Sec. III, we show the generalizations of the Fourier law and the heat fluctuation theorem, 
		and discuss the zeroth law of thermodynamics for athermal systems. 
		In Sec. IV, we present derivations of our main results. 
		In Sec. V, we conclude this paper with some remarks. 
		In Appendix A, we discuss the cumulant functional and the $n$-points delta functions. 
		In Appendix B, we review the formulation of the $\ast$ integral. 
		In Appendix C, we show a detailed analysis for a weakly quartic potential. 
		In Appendix D, we numerically show non-linear effect in the generalized heat fluctuation theorem. 
		
\section{Model}
		\begin{figure}
			\centering
			\includegraphics[width=70mm,clip]{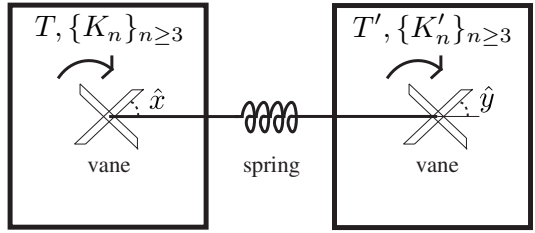}
			\caption{
						A schematic picture of heat conduction between athermal environments. 
						The noise from the left (right) environment is characterized by the noise intensity $T$ ($T'$) and the higher order cumulants $\{K_{n}\}_{n\geq3}$ ($\{K'_{n}\}_{n\geq3}$). 
					}
		\end{figure}
		We consider a non-Gaussian stochastic model of a Brownian motor which consists of two vanes that are attached to two environments which can be athermal by a spring (see Fig. 1).
		The vanes are driven by athermal fluctuations in the environments, and the spring conducts energy current induced by the fluctuations.  
		We refer to the energy current as the heat current. 
		The motion of the vanes is described by the following Langevin equations: 
		\begin{equation}
			\frac{d\hat x}{dt}	=-\frac{\partial U(\hat x-\hat y)}{\partial \hat x}+\hat \xi, \>\>\>
			\frac{d\hat y}{dt}	=-\frac{\partial U(\hat x-\hat y)}{\partial \hat y}-\hat \eta,\label{eq:2}
		\end{equation}
		where $\hat{x},\hat{y}$ are the angles of the vanes, 
		$U(\hat{x}-\hat{y})$ is the dimensionless potential energy of the spring, 
		and $\hat{\xi},\hat{\eta}$ are independent white non-Gaussian noises that characterize the fluctuations from the athermal environments. 
		In the following, $\langle\hat{A}\rangle$ denotes the ensemble average of a stochastic variable $\hat{A}$, and the Boltzmann constant is taken to be unity. 
		The cumulants of the noises are given by 
		\begin{align}
			\langle\hat\xi(t)\rangle=\langle&\hat\eta(t)\rangle=0, \\
			\langle\hat\xi(t_1)\hat\xi(t_2)\rangle_c&=2T\delta(t_1-t_2),\\
			\langle\hat\eta(t_1)\hat\eta(t_2)\rangle_c&=2T'\delta(t_1-t_2),\\
			\langle\hat\xi(t_1)\hat\xi(t_2)\dots\hat\xi(t_{n})\rangle_c&=K_{n}\delta_{n}(t_1,t_2,\dots,t_{n}), \\
			\langle\hat\eta(t_1)\hat\eta(t_2)\dots\hat\eta(t_{n})\rangle_c&=K'_{n}\delta_{n}(t_1,t_2,\dots,t_{n}), 
		\end{align}
		where $\langle\hat\xi(t_1)\dots\hat\xi(t_n)\rangle_c$ denotes the $n$-th cumulant,
		and $\delta_n(t_1,\dots,t_n)$ is a $n$-point delta function with an positive integer $n$ 
		(see Appendix A for details). 
		We write $K_2\equiv2T$ and $K_2'\equiv2T'$.
		On the basis of stochastic energetics~\cite{Seifert4,Sekimoto1,Kanazawa,Sekimoto2,Sekimoto3}, the heat current is defined by 
		\begin{equation}\label{eq:stochastic_heat}
			\frac{d\hat Q }{dt} = \left(- \frac{d\hat x}{dt} + \hat \xi  \right)\ast \frac{d\hat x}{dt}, 
		\end{equation}
		where $\ast$ describes a stochastic integral that is defined as a white noise limit of a colored noise. 
		As discussed in Ref.~\cite{Kanazawa} and Appendix B in detail, 
		the $\ast$ integral for an arbitrary function $f(\hat{x}(s))$ is given by 
		\begin{equation}
			\int_0^t ds \hat \xi(s)\ast f(\hat x(s)) \equiv 
			\lim_{\varepsilon \rightarrow +0}\lim_{\Delta t\rightarrow +0}\sum_{i=0}^{N-1} \Delta t \hat \xi_\varepsilon(t_i) f(\hat x(t_i)), 
		\end{equation} 
		where $\Delta t\equiv t/N$, $t_i \equiv i\Delta t$, 
		and $\hat{\xi}_\varepsilon(t_i)$ is the colored noise with a correlation time $\varepsilon>0$ such that $\lim_{\varepsilon\to0}\hat{\xi}_\varepsilon(t_i)=\hat{\xi}(t_i)$. 
		For non-Gaussian noises, the definition of the heat is not consistent with the Stratonovich integral but with the $\ast$ integral~\cite{Kanazawa}. 
		We note that the Stratonovich and $\ast$ integrals are the same for Gaussian noises. 
		In the Gaussian case with $K_{n}=K'_{n}=0$ for $n\geq3$, we can show that the motor obeys the conventional Fourier law and the heat fluctuation theorem: 
		\begin{equation}
			J = -\mathcal{\kappa}\Delta T, \label{eq:ordinary_Fourier's_law}
		\end{equation}
		\begin{equation}
			\lim_{t\rightarrow \infty}\frac{1}{t}\ln{\frac{P(+q,t)}{P(-q,t)}} =\Delta \beta q, \label{eq:ordinary_FT}
		\end{equation}
		where $J=\langle{}d\hat{Q}/dt\rangle_{\rm{SS}}\equiv\lim_{t\rightarrow\infty}\langle{}d\hat{Q}/dt\rangle$ is the average heat current in the steady state, 
		$\kappa$ is the thermal conductivity, 
		$\Delta T\equiv T'-T$, 
		$\Delta\beta\equiv 1/T'-1/T$, 
		$P(q,t)\equiv P(\hat{Q}(t)=qt)$,
		and $q$ is the time average heat current. 
		
\section{Main results}
	In this section, we summarize the main results in this paper. The derivation of them will be presented in Sec. IV associated with Appendices.
	\subsection{Generalized Fourier Law}
		We now discuss the generalized Fourier law for an arbitrary potential $U(\hat{x}-\hat{y})$ on the basis of the perturbation in terms of $\Delta T$, $K_{n}$ and $K'_{n}$ with $n\geq3$. 
		In the first order perturbation, 
		we obtain the generalized Fourier law:
		\begin{align}
			J &= - \sum_{n=2}^\infty \kappa_{n}\Delta K_n, \label{eq:heat_flux_general}\\
			\kappa_n &= \frac{1}{2\cdot n!} \left<\frac{d^nU(\hat z)}{d\hat z^n}\right>_{\mathrm{eq}},
		\end{align}
		where $\hat{z}\equiv\hat{x}-\hat{y}$, $\Delta{}K_n\equiv{}K'_n-K_n$, 
		and 
		\begin{equation}
			\langle f(\hat{z})\rangle_{\mathrm{eq}}\equiv\int_{-\infty}^{\infty}dzf(z)P_{\mathrm{eq}}(z)
		\end{equation} for an arbitrary function $f(\hat{z})$ with 
		\begin{equation}
			P_{\mathrm{eq}}(z)\equiv \frac{e^{-U(z)/T}}{\int_{-\infty}^{\infty}dye^{-U(y)/T}}.
		\end{equation}
		This is the first main result of this paper. 
		The first term on the right hand side (rhs) of Eq.~{(\ref{eq:heat_flux_general})} , i.e. $-2\kappa_2\Delta T$, corresponds to the conventional Fourier law, 
		and the other terms describe the correction terms due to the non-Gaussianity of the noises. 
		This result implies that the heat is conducted from the environment with the higher non-Gaussianity to the other environment. 
		Particularly in the case of $T=T'$, the Gaussian term of the rhs of Eq.~{(\ref{eq:heat_flux_general})} vanishes, but non-Gaussian terms drive the heat current.  
		We note that the effect of the $n$-th cumulant is induced by the $n$-th differential coefficient of the potential, which implies that the quartic potential model is minimum to reveal the non-Gaussian effects. 
		In fact, if the potential is harmonic, the non-Gaussian effects vanish in Eq.~{(\ref{eq:heat_flux_general})}. 

		\begin{figure}
			\centering
			\includegraphics[width=80mm]{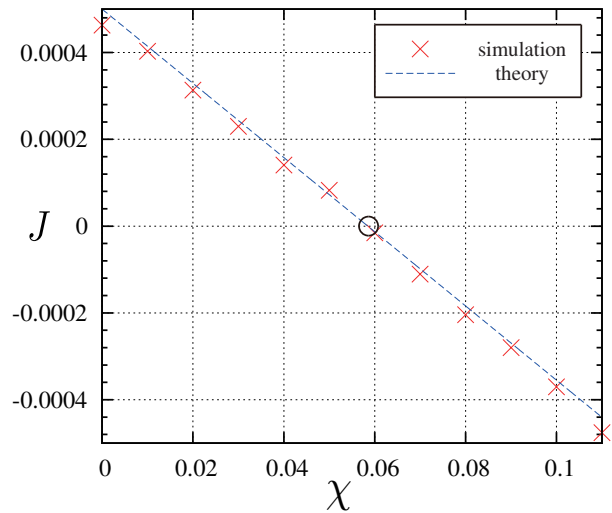}
			\caption{	(Color online)
						Numerical verification of Eq.~{(\ref{eq:heat_flux_general})}, where $\chi$ characterizes the nonlinearity of the potential. 
						The dashed line is theoretically obtained from Eq.~{(\ref{eq:heat_flux_general})}, 
						the cross points show the numerical data of our Monte Carlo simulation, 
						and the open circle indicates the point at which the direction of the heat current is switched. 
						As $\chi$ becomes larger, the heat current becomes smaller. 
						We assume the ergodicity
						$\langle d\hat{Q}/dt\rangle_{\rm{SS}}=\lim_{T\rightarrow\infty}\left[1/T\int_0^Tds\left(d\hat{Q}/dt\right)\right]$ 
						and calculated the long time average instead of the ensemble average. 
						The time step is given by $1.0\times10^{-4}$ and the entire time interval for the average is $1.0\times10^9$. 
					}
		\end{figure}
		We have numerically verified Eq.~{(\ref{eq:heat_flux_general}) for a quartic potential $U(\hat{z})=\hat{z}^2/2+\chi\hat{z}^4/4$ with $\chi>0$. 
		For simplicity, we assume that $\hat\xi(t)$ is a white Gaussian noise 
		and that $\hat\eta(t)$ is a two-sided Poisson noise with intensity $\sqrt{2T'/\lambda'}$ and transition rate $\lambda'/2$: $\hat{\eta}(t)=\sum_{i}\sqrt{2T'/\lambda'}\delta(t-\hat{t_i})+\sum_{i}(-\sqrt{2T'/\lambda'})\delta(t-\hat{s_i})$, 
		where $\hat{t_i},\hat{s_i}$ are times at which Poisson flights happen. 
		Figure 2 shows our numerical results with $T=0.300$, $T'=0.299$, and $\lambda'=5.0$. 
		We plot the average heat current by changing $\chi$. 
		The direction of the heat current is changed at $\chi\simeq0.058$, which implies that the direction of the heat current depends on the potential profile of the heat conductor. 
		We explicitly present the detailed analysis for weakly quartic case in Appendix C. 
		
	\subsection{Generalized heat fluctuation theorem}
		We next discuss a correction term to the conventional heat fluctuation theorem on the basis of the perturbation in terms of $K_{n}$ and $K'_{n}$ with $n\geq3$. 
		Here we do not assume that $\Delta T$ is also small.
		For simplicity, we consider the case of a harmonic potential with $U(\hat{z})=\hat{z}^2/2$. 
		We obtain a correction term to the heat fluctuation theorem up to the first order perturbation: 
		\begin{align}\label{eq:Modified_FT}
			&\lim_{t\rightarrow \infty}\frac{1}{t}\ln{\frac{P(+q,t)}{P(-q,t)}} 
			= \Delta \beta q + \sum_{n=2}^\infty \left[K_{2n}\Xi_{2n}(q) + K'_{2n}\Xi'_{2n}(q)\right],\\
			&\Xi _{2n}(q) \equiv \mathrm{As}\left[\frac{1}{(4T ^2)^nn!}\left(+q+\frac{2q^2 +T \Delta T}{2\sqrt{q^2+TT'}} \right)^n\right],\\ 
			&\Xi'_{2n}(q) \equiv \mathrm{As}\left[\frac{1}{(4T'^2)^nn!}\left(-q+\frac{2q^2 -T'\Delta T}{2\sqrt{q^2+TT'}} \right)^n\right],
		\end{align}
		where $\mathrm{As}[f(q)]\equiv f(q)-f(-q)$ is the antisymmetric part of an arbitrary function $f(q)$. 
		This is the second main result of this paper. 
		Although the conventional Fourier law~{(\ref{eq:ordinary_Fourier's_law})} holds for a harmonic potential, 
		the conventional heat fluctuation theorem~{(\ref{eq:ordinary_FT})} should be modified as Eq.~{(\ref{eq:Modified_FT})} even for the harmonic potential.
		This implies that the effect of the non-Gaussianity appears only in the higher cumulants in this case. 

		Let us consider a special case where $\hat\xi$ and $\hat\eta$ are the two-sided Poisson noises 
		with intensities $\sqrt{2T/\lambda}$ and $\sqrt{2T'/\lambda'}$ and transition rates $\lambda/2$ and $\lambda'/2$, respectively. 
		In this case, Eq.~{(\ref{eq:Modified_FT})} reduces to a simpler form:
		\begin{equation}
			\lim_{t\rightarrow \infty}\frac{1}{t}\ln{\frac{P(+q,t)}{P(-q,t)}}=
			2\Delta \beta q 
			+ 2\lambda e^{\frac{2q^2+T\Delta T}{4T\lambda \sqrt{q^2+TT'}}}\sinh{\frac{q}{2T\lambda}} - 2\lambda'e^{\frac{2q^2-T'\Delta T}{4T'\lambda'\sqrt{q^2+TT'}}}\sinh{\frac{q}{2T'\lambda'}}.\label{eq:g_FT_P}
		\end{equation}
		We note that the Gaussian limit is given by $\lambda,\lambda'\rightarrow\infty$. 
		Particularly, let us focus on the case of $T=T'$ and $\lambda'=\infty$, where $J=0$ holds. 
		We note that $\hat\eta$ is the Gaussian noise in the limit $\lambda'\rightarrow\infty$. 
		In this case, the fluctuation function $F(q)\equiv\lim_{t\rightarrow\infty}(1/t)\ln{P(+q,t)/P(-q,t)}$ is positive for $q>0$, 
		which is interpreted as follows: 
		although $P(q,t)$ converges to $\delta(q)$ in the limit $t\rightarrow\infty$, the convergence speed is asymmetric in terms of $q$.
		Thus, the heat tends to flow from the environment with the higher non-Gaussianity to the other environment, though the average heat flux is zero. 
		
		\begin{figure}
			\centering
			\includegraphics[width=80mm]{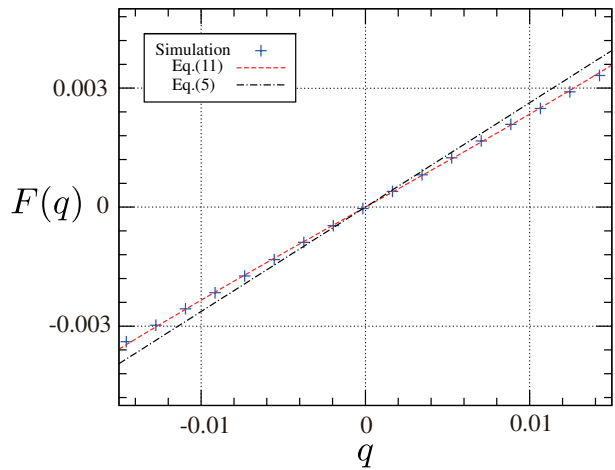}
			\caption{	
						(Color online)
						Numerical verification of Eq.~{(\ref{eq:g_FT_P})}. 
						The red broken line is obtained from Eq.~{(\ref{eq:g_FT_P})}, 
						the black chain line is obtained from Eq.~{(\ref{eq:ordinary_FT})}, 
						and the blue cross points show the numerical data of our simulation. 
						We perform the Monte Carlo simulation to make the histogram of the heat distribution function, 
						and numerically obtain the fluctuation function $F(q)$. 
						The bin-width for the heat histogram is $0.03$, 
						the time step is $0.0001$, and 
						the number of samples is $5\times10^7$. 
					}
		\end{figure}
		We have numerically checked the validity of Eq.~{(\ref{eq:g_FT_P})} as shown in Fig. 3. 
		By taking $t=1000$, $T=0.20$, $T'=0,19$, $\lambda=2.0$, and $\lambda'=\infty$,
		we numerically obtain the fluctuation function $F(q)$ 
		and compare it with Eqs.~{(\ref{eq:g_FT_P})} and {(\ref{eq:ordinary_FT})}. 
		We observe a significant deviation from the conventional heat fluctuation theorem~{(\ref{eq:ordinary_FT})}, 
		and the deviation is consistent with our result~{(\ref{eq:g_FT_P})}. 
		We also demonstrate how the heat fluctuation theorem is modified for the case that the nonlinear correction is relevant in Appendix D. 

	\subsection{A generalized zeroth law of thermodynamics}
		We next discuss the zeroth law of thermodynamics~\cite{Casas,SST,Hatano,Ren}. 
		In the case of non-Gaussian noises with $\Delta K_{n}\neq0$ for $n\geq3$, 
		Eq.~{(\ref{eq:heat_flux_general})} implies that the condition of $J=0$ explicitly depends on the spring potential $U(\hat{z})$. 
		In contrast, in the case of Gaussian noises, $J=0$ if and only if $\Delta T=0$. 
		Therefore, the zeroth law of thermodynamics is not universally valid for non-Gaussian noises; the condition of $J=0$ depends on the details of the contact device (i.e., the spring).
		When we fix the contact device, however, there is a transitive relation for thermal equilibrium 
		and we can introduce an indicator characterizing the direction of heat current. 
		
		To show this, we consider three athermal environments, AE$^{(1)}$, AE$^{(2)}$ and AE$^{(3)}$, 
		whose fluctuations are characterized by the cumulants $(T^{(i)},\{K_{n}^{(i)}\}_{n\geq3})$ with $i=1,2,3$. 
		If we link the contact device between AE$^{(i)}$ and AE$^{(j)}$ $(i,j=1,2,3,\>i\neq j)$, 
		the average heat current between them is given by $J^{(ij)}=-\sum_{n\geq 2}\kappa_n\Delta^{(ij)}K_n$, 
		where $\Delta^{(ij)}K_n\equiv K_n^{(j)}-K_n^{(i)}$. 
		We can then show the transitive relation: if $J^{(12)}=0$ and $J^{(23)}=0$, then $J^{(13)}=0$. 
		We also introduce a device-dependent indicator $\mu_{U}(T,\{K_n\}_{n\geq3})$ as 
		\begin{equation}
			\mu_{U}(T,\{K_n\}_{n\geq 3}) = \sum_{n=2}^\infty \kappa_n K_n,
		\end{equation}
		which characterizes the direction of the average heat current 
		\begin{equation}
			J^{(ij)}=\mu_U(T^{(i)},\{K_{n}^{(i)}\}_{n\geq3})-\mu_U(T^{(j)},\{K_{n}^{(j)}\}_{n\geq3}). 
		\end{equation}
		In this sense, $\mu_U(T^{(i)},\{K_{n}^{(i)}\}_{n\geq3})$ plays a corresponding role to that of the temperature in equilibrium thermodynamics. 
		Such a device-dependent temperature has also been introduced in Refs.~\cite{Seifert1,Seifert2} for driven lattice gases, which implies that our results may hold beyond our model.
				
\section{Derivations}
	In this section, we present the details of the derivation of the main results introduced in the previous section. 
	This section consists of two parts: the derivations of the generalized Fourier law and of the generalized heat fluctuation theorem.
	
	\subsection{Derivation of the generalized Fourier law (\ref{eq:heat_flux_general})}
		We now present the derivation of the generalized Fourier law (\ref{eq:heat_flux_general}).
		By introducing a new variable $\hat{z}\equiv \hat{x}-\hat{y}$, 
		Eqs.~{(\ref{eq:2}) reduce to a single equation: 
		\begin{equation}\label{eq:single_eq}
			\frac{d\hat z}{dt} = -2\frac{dU(\hat z)}{d \hat z} + \hat \xi + \hat \eta.
		\end{equation}
		Let us introduce a stochastic distribution function as $\hat{\mathcal{P}}(z,t)\equiv \delta(z-\hat z(t))$,
		which satisfies the stochastic Liouville equation~\cite{Kubo2} 
		\begin{align}
			\frac{\partial \hat{\mathcal{P}}(z,t)}{\partial t} 	&= -\frac{\partial }{\partial z}\left(\frac{d\hat z}{dt} \ast \hat{\mathcal{P}}(z,t)\right)\notag\\
															&= 2\frac{\partial }{\partial z}\frac{dU}{dz}\hat{\mathcal{P}}(z,t) - \frac{\partial }{\partial z}(\hat \xi +\hat \eta) \ast \hat{\mathcal{P}}(z,t).
		\end{align}
		Using the transformation formulas from the $\ast$ integral to the It\^o one [i.e., Eqs.~{(\ref{eq:transform1})} and~{(\ref{eq:transform2})}], 
		we obtain the master equation of the distribution function $P(z,t)\equiv \langle \hat{\mathcal{P}}(z,t)\rangle$ as 
		\begin{align}
			\frac{\partial P}{\partial t} 	&= 
			\left[2\frac{\partial }{\partial z}\frac{dU}{dz}+(T+T')\frac{\partial^2 }{\partial z^2} + \sum_{n=3}^{\infty}\frac{K_n+K'_n}{(-1)^nn!}\frac{\partial^n }{\partial z^n}\right]P \notag\\
											&= 2\frac{\partial }{\partial z}(L_0+L_1)P(z,t),\label{eq:Kramers_Moyal_1}
		\end{align}
		where 
		\begin{equation}
			L_0 \equiv \frac{dU}{dz} + T\frac{\partial}{\partial z}, \>\>\>
			L_1 \equiv \frac{\Delta T}{2}\frac{\partial}{\partial z} + \sum_{n=3}^\infty \frac{K_{n}+K'_{n}}{(-1)^n2\cdot n!}\frac{\partial^{n-1}}{\partial z^{n-1}}. \notag
		\end{equation} 
		The normalization of the probability is given by $\int_{-\infty}^{\infty}dy P(y,t) = 1$.
		The steady solution of Eq.~{(\ref{eq:Kramers_Moyal_1})} satisfies the relation $(L_0+L_1)P_{\mathrm{SS}}(z)=0$, where $P_{{\mathrm{SS}}}(z) \equiv \lim_{t\rightarrow \infty}P(z,t)$.
		We assume that $\Delta T$, $K_n$ and $K'_n$ are perturbative terms. 
		In the first order perturbation, we expand the steady solution of as $P_{{\mathrm{SS}}}(z) = P_0(z) + P_1(z)$, where $P_0$ and $P_1$ are the unperturbative and perturbative steady distributions, respectively. 
		Here, $P_0(z)$ and $P_1(z)$, respectively, satisfy the equations
		\begin{align}
			&T\frac{dP_0(z)}{dz}+\frac{dU}{dz}P_0(z) =0,\\
			&T\frac{dP_1(z)}{dz}+\frac{dU}{dz}P_1(z) =-\frac{\Delta T}{2}\frac{dP_0(z)}{dz} - \sum_{n=3}^\infty \frac{(-1)^n}{2}\frac{K_{n}+K'_{n}}{n!}\frac{d^{n-1}P_0(z)}{dz^{n-1}}.
		\end{align}
		Thus we obtain the solutions
		\begin{align}
			&P_0(z) = \frac{e^{-U(z)/T}}{\int_{-\infty}^{\infty}dy e^{-U(z)/T}} =P_{\mathrm{eq}}(z),\\
			&P_1(z)	= P_0(z) \biggl[ C + \frac{\Delta T U(z)}{2T^2} - \sum_{n=3}^\infty \frac{(-1)^n}{2T}\frac{K_n+K'_n}{n!}\int_{0}^{z} dyP_0^{-1}(y)\frac{d^{n-1}P_0(y)}{dy^{n-1}}\biggr], \label{eq:solution_1}
		\end{align}
		where $C$ is a renormalization constant determined by $\int_{-\infty}^{\infty} dy P_1(y) = 0$. 
		The average heat flux is given by 
		\begin{align}
			J 	\equiv& \left<\left(-\frac{d\hat x}{dt} + \hat \xi \right)\ast \frac{d\hat x}{dt}\right>_{\mathrm{SS}} \notag\\
				=& \left<- \left(\frac{dU}{d\hat z}\right)^2 + \frac{dU}{d\hat z}\ast \hat \xi \right>_{\mathrm{SS}} \notag\\
				=&\left<T\frac{d^2U}{d\hat z^2} - \left(\frac{dU}{d\hat z}\right)^2\right>_{\mathrm{SS}} +\sum_{n=3}^\infty \frac{K_{n}}{n!}\left<\frac{d^{n} U}{d\hat z^{n}}\right>_{\mathrm{SS}}\notag\\
				=&\int_{-\infty}^\infty dz \left(P_0(z)+P_1(z)\right)\left[T\frac{dU}{dz^2} - \left(\frac{dU}{dz}\right)^2\right] + \sum_{n=3}^\infty \frac{K_n}{n!}\left<\frac{d^n U}{d\hat z^n}\right>_{\mathrm{eq}},
		\end{align}
		where we have used Eq.~{(\ref{eq:transform1})}, 
		$\langle \hat A \rangle_{\mathrm{SS}} \equiv \int_{-\infty}^\infty dzP_{\mathrm{SS}}(z)A(z)$, and $\langle \hat A\rangle_{\mathrm{eq}} \equiv \int_{-\infty}^\infty dzP_0(z)A(z)$. 
		Using the following equalities
		\begin{align}
			&T\frac{d^2U}{dz^2} - \left(\frac{dU}{dz}\right)^2 = -\frac{T^2}{P_0(z)}\frac{d^2 P_0(z)}{dz^2}, \notag\\
			\int_{-\infty}^\infty dz P_0(z)&\frac{d}{dz} \left(P^{-1}_0(z)\frac{d^{n-1}P_0(z)}{dz^{n-1}}\right) 
			= \left<\frac{(-1)^{n-1}}{T}\frac{d^{n}U}{d\hat z^{n}}\right>_{\mathrm{eq}},\notag
		\end{align}
		we obtain 
		\begin{equation}
			\int_{-\infty}^\infty dz P_0(z)\left[T\frac{dU}{dz^2} - \left(\frac{dU}{dz}\right)^2\right] =0,
		\end{equation}
		and
		\begin{align}
			&\int_{-\infty}^\infty dz P_1(z)\left[T\frac{dU}{dz^2} - \left(\frac{dU}{dz}\right)^2\right]\notag\\
			&=-T^2\int_{-\infty}^\infty dzP_0(z) \frac{d^2}{dz^2} \left(\frac{P_1(z)}{P_0(z)}\right)\notag\\
			&=-\frac{\Delta T}{2}\left<\frac{d^2U}{d\hat z^2}\right>_{\mathrm{eq}} - \sum_{n=3}^\infty \frac{K_n+K'_n}{2\cdot n!} \left<\frac{d^nU}{d\hat z^n}\right>_{\mathrm{eq}}.
		\end{align}
		Thus we obtain
		\begin{equation}
			J = -\frac{\Delta T}{2}\left<\frac{d^2 U}{d\hat z^2}\right>_{\mathrm{eq}} - \sum_{n=3}^{\infty}\frac{K'_n-K_n}{2\cdot n!}\left<\frac{d^nU}{d\hat z^n}\right>_{\mathrm{eq}},
		\end{equation}
		which is the generalized Fourier law~{(\ref{eq:heat_flux_general})}.
		
	\subsection{Derivation of the generalized heat fluctuation theorem (\ref{eq:Modified_FT})}
		
		We here derive the generalized heat fluctuation theorem~{(\ref{eq:Modified_FT})}. 
		We first assume a harmonic potential $U(\hat z) = \hat z^2/2$ and derive the master equation of a distribution function for $\hat z$ and $\hat Q$. 
		Let us introduce a stochastic distribution function $\hat{\mathcal{P}}(z,Q,t) \equiv \delta (z-\hat z(t))\delta (Q-\hat Q(t))$. 
		The stochastic Liouville equation for $\hat{\mathcal{P}}(z,Q,t)$ is given by 
		\begin{align}
			\frac{\partial \hat{\mathcal{P}}(z,Q,t)}{\partial t} 
				&= -\left[\frac{\partial }{\partial z}\!\frac{d\hat z}{dt} +\frac{\partial }{\partial Q}\frac{d\hat Q}{dt}\right]\ast\hat{\mathcal{P}}(z,Q,t)\notag\\
				&= 2\frac{\partial }{\partial z}\left[z \hat{\mathcal{P}}(z,Q,t)\right] + \frac{\partial }{\partial Q}\left[z^2 \hat{\mathcal{P}}(z,Q,t)\right] 
					-\frac{\partial }{\partial z}(\hat \xi+\hat \eta) \ast \hat{\mathcal{P}}(z,Q,t) 
					-\frac{\partial }{\partial Q}z \hat{\mathcal{P}}(z,Q,t)\ast \hat \xi.
		\end{align}
		Using Eqs.~{(\ref{eq:transform1})} and~{(\ref{eq:transform2})},
		we obtain the master equation of $P(z,Q,t)\equiv \langle  \hat{\mathcal{P}}(z,Q,t)\rangle $ as 
		\begin{equation}
			\frac{\partial P(z,Q,t)}{\partial t} 	= \biggl[2\frac{\partial }{\partial z}z + \frac{\partial }{\partial Q}z^2 
				+T\left(\frac{\partial }{\partial z}+z\frac{\partial }{\partial Q}\right)^2
				+T'\frac{\partial^2 }{\partial z^2} 
				+\sum_{n=3}^\infty \frac{(-1)^n}{n!} \left\{K_n\left(\frac{\partial }{\partial z}+z\frac{\partial}{\partial Q}\right)^n+K'_n\left(\frac{\partial}{\partial z}\right)^n\right\} \biggr] P(z,Q,t).
		\end{equation}
		By introducing the Laplace transformation of $P(z,Q,t)$ as $\rho_v(z,t) \equiv \int_{-\infty}^\infty dQe^{-vQ}P(z,Q,t)$, 
		we derive the modified master equation for $\rho_v(z,t)$ as 
		\begin{align}
			&\frac{\partial \rho_v(z,t)}{\partial t} = \left(L_0^{v} + L_1^{v}\right)\rho_v(z,t),\label{eq:eigen_equation}\\
			&L_0^v \equiv 2\frac{\partial }{\partial z}z + vz^2 + T\left(\frac{\partial }{\partial z}+zv\right)^2 + T'\frac{\partial^2}{\partial z^2},\notag \\
			&L_1^v \equiv \sum_{n=3}^\infty \frac{(-1)^n}{n!}\left\{K_n\left(\frac{\partial }{\partial z}+zv\right)^n +K'_n\left(\frac{\partial }{\partial z}\right)^n\right\},\notag
		\end{align}
		The adjoint operators of $L_0^v$ and $L_1^v$ are respectively given by
		\begin{align}
			(L_0^v)^\dagger &\equiv -2z\frac{\partial}{\partial z} + vz^2 +T\left(-\frac{\partial}{\partial z}+zv\right)^2 + T'\frac{\partial^2}{\partial z^2}, \notag\\
			(L_1^v)^\dagger &\equiv \sum_{n=3}^\infty \frac{(-1)^n}{n!}\left\{K_n\left(-\frac{\partial }{\partial z}+zv\right)^n +K'_n\left(-\frac{\partial }{\partial z}\right)^n\right\}.\notag
		\end{align}
		Let us denote an eigenfunction of the operator $L_0^v+L_1^v$ by $\psi_n^v(z) \>\>\>(n=0,1,2,\dots)$ and the corresponding eigenvalue by $\mu_n^v \>\>\>(n=0,1,2,\dots)$. 
		We assume that the eigenvalues satisfy $\mathrm{Re}(\mu_n^v) \leq \mathrm{Re}(\mu_m^v)$ for $n > m$, where $\mathrm{Re}(a)$ is the real part of an arbitrary complex number $a$. 
		We denote an eigenfunction of the operator $(L_0^v)^\dagger+(L_1^v)^\dagger$ by $\phi_n^v(z) \>\>\>(n=0,1,2,\dots)$ and the corresponding eigenvalue by $\nu_n^v \>\>\>(n=0,1,2,\dots)$. 
		According to the Perron-Frobenius theory~\cite{Nemoto}, we can generally set $\nu_n^v = (\mu_n^v)^*$ for any $n$ and the largest eigenvalues $\nu_0^v$ and $\mu_0^v$ are real. 
		Furthermore, the largest eigenvalue $\mu_0^v$ is known to be equal to the scaled cumulant generating function~\cite{Nemoto}
		\begin{equation}
			\Psi(v) \equiv \lim_{t\rightarrow \infty} \frac{1}{t}\ln{\langle e^{-v\hat Q(t)}\rangle}
		\end{equation}.  
		The orthonormal conditions for the eigenfunctions are given by 
		\begin{equation}
			\int_{-\infty}^\infty dy (\phi_n(y))^*\psi_m(y) = \delta_{n,m},\label{eq:orthonormal}
		\end{equation}
		where $n$ and $m$ are non-negative integers and $\delta_{n,m}$ is the Kronecker delta.
		To solve this eigenvalue problem, we perform a perturbative calculation in terms of $K_{n}$ and $K'_{n}$ ($n\geq 3$). 
		We expand the largest eigenvalue $\mu_0^v$ and the corresponding eigenfunctions $\psi_0^v(z), \phi_0^v(z)$ as  
		\begin{align}
			\mu_0^v &= \mu_{0,0}^v + \mu_{0,1}^v, \\
			\psi_0^v(z) &= \psi_{0,0}^v(z) + \psi_{0,1}^v(z), \\
			\phi_0^v(z) &= \phi_{0,0}^v(z) + \phi_{0,1}^v(z), 
		\end{align}
		where $\mu_{0,0}(v)$, $\psi_{0,0}^v(z)$, and $\phi_{0,0}^v(z)$ are the unperturbative terms, and $\mu_{0,1}(v)$, $\psi_{0,1}^v(z)$, and $\phi_{0,1}^v(z)$ are the perturbative terms. 
		In the first order perturbation, we obtain 
		\begin{align}
			L_0^v\psi_{0,0}^v(z) &= \mu_{0,0}^v\psi_{0,0}^v(z), \label{eq:eigen_1}\\
			(L_0^v)^\dagger\phi_{0,0}^v(z) &= \mu_{0,0}^v\phi_{0,0}^v(z),\label{eq:eigen_2}\\
			L_0^v\psi_{0,1}^v(z) + L_1^v\psi_{0,0}^v(v)&= \mu_{0,0}^v \psi_{0,1}^v(z)+ \mu_{0,1}^v\psi_{0,0}^v(z)\label{eq:eigen_3}.
		\end{align}
		The solutions of Eqs. (\ref{eq:eigen_1}) and (\ref{eq:eigen_2}) are given by~\cite{Visco,Wijland}
		\begin{align}
			\mu_{0,0}^v 	&= 1 - \sqrt{(1+Tv)(1-T'v)}, \\
			\psi_{0,0}^v(z) &= \exp{\left(-\frac{z^2}{2T_1^*}\right)}, \label{eq:eigen_solution_2}\\
			\phi_{0,0}^v(z)	&= \sqrt{\frac{\frac{1}{T_1^*}+\frac{1}{T_2^*}}{2\pi}}\exp{\left(-\frac{z^2}{2T_2^*}\right)}, \label{eq:eigen_solution}
		\end{align}
		where $T_1^* \equiv (T+T')/(\sqrt{(1-T'v)(1+Tv)}+1+Tv)$ and $T_2^* \equiv (T+T')/(\sqrt{(1-T'v)(1+Tv)}-1-Tv)$. 
		Multiplying $\phi_{0,0}^v(z)$ to the both sides of Eq.~{(\ref{eq:eigen_3})} and integrating them by $z$, we obtain 
		\begin{widetext}
			\begin{align}
				\mu_{0,1}^v &= \int_{-\infty}^\infty dz \phi_{0,0}^v (z)L_1^v \psi_{0,0}^v(z)\notag\\
				&= \sqrt{\frac{\frac{1}{T_1^*}+\frac{1}{T_2^*}}{2\pi}}\left[
					\sum_{n=3}^\infty \frac{(-1)^nK_n}{n!} \int_{-\infty}^\infty dz e^{-\frac{z^2}{2}\left(\frac{1}{T_2^*}+v\right)}\frac{d^{n}}{dz^n}e^{-\frac{z^2}{2}\left(\frac{1}{T_1^*}-v\right)}
					+ \sum_{n=3}^\infty \frac{(-1)^nK'_n}{n!}\int_{-\infty}^\infty dz e^{-\frac{z^2}{2T_2^*}}\frac{d^{n}}{dz^n}e^{-\frac{z^2}{2T_1^*}}\right]\notag\\
				&=\sum_{n=2}^\infty \frac{K_{2n}}{n!}\left[\frac{-v}{4}\sqrt{\frac{1-T'v}{1+Tv}}\right]^n + \sum_{n=2}^\infty \frac{K'_{2n}}{n!}\left[\frac{v}{4}\sqrt{\frac{1+Tv}{1-T'v}}\right]^n,
			\end{align}
		\end{widetext}
		where we have used Eqs. {(\ref{eq:orthonormal})}, {(\ref{eq:eigen_2})}, {(\ref{eq:eigen_solution_2})}, {(\ref{eq:eigen_solution})}, and identities for Hermite polynomial $H_n(z)$:
		\begin{align}
			&\frac{d}{dz} +vz =e^{-\frac{vz^2}{2}}\frac{d}{dz}e^{\frac{vz^2}{2}},\>\>\>
			H_n(z) \equiv (-1)^n e^{z^2}\frac{d^n}{dz^n}(e^{-z^2}),\notag\\
			&\int_{-\infty}^\infty dze^{-\frac{z^2}{2\alpha}}H_{n}(z) = 	\begin{cases}
																			\sqrt{2\pi\alpha}\frac{n!}{(n/2)!}(2\alpha-1)^{n/2} & $(even $n$)$ \cr
																			0 & $(odd $n$)$
																		\end{cases}.\notag
		\end{align}
		Thus, we obtain the scaled cumulant generating function 
		\begin{align}
			\Psi(v)   &= \Psi_0(v) + \Psi_1(v), \\
			\Psi_0(v) &\equiv 1 - \sqrt{(1-T'v)(1+Tv)}, \notag\\
			\Psi_1(v) &\equiv \sum_{n=2}^\infty \frac{K_{2n}}{n!}\left[\frac{-v}{4}\sqrt{\frac{1-T'v}{1+Tv}}\right]^n + \sum_{n=2}^\infty \frac{K'_{2n}}{n!}\left[\frac{v}{4}\sqrt{\frac{1+Tv}{1-T'v}}\right]^n.\notag
		\end{align}
		We note that the scaled cumulant generating function has singular points $v = -1/T, 1/T'$, near which the perturbation is not valid. 
		
		The asymptotic form of the distribution function $P(q,t)=\langle \delta(q-\hat Q(t)/t)\rangle$ is related to the cumulant generating function~\cite{Touchette2} as
		\begin{equation}
			\lim_{t\rightarrow \infty}\frac{1}{t}\ln{P(q,t)} = v^*q + \Psi(v^*),
		\end{equation}
		where $v=v^*$ is the point at which $vq + \Psi(v)$ is minimum. 
		The explicit form of $v^*$ is given by the condition $q + d\Psi(v)/dv|_{v=v^*} = 0$.
		In the first order perturbation, we obtain 
		\begin{align}
			\lim_{t\rightarrow \infty}\frac{1}{t}\ln{P(q,t)} &= v_0^*q + \Psi_0(v_0^*) + \Psi_1(v_0^*),\label{eq:large_deviation}\\
			&q + \frac{d\Psi_0(v)}{dv}\bigg|_{v=v_0^*} = 0,\label{eq:saddle_point}
		\end{align}
		where we have expanded $v=v_0^*+ v_1^*$ with the unperturbative and perturbative terms $v_0^*$ and $v_1^*$, respectively. 
		By solving Eq.~{(\ref{eq:saddle_point})}, $v_0^*$ is explicitly written as 
		\begin{equation}
			v_0^* = \frac{\Delta \beta}{2} - \frac{(\beta +\beta')q}{2\sqrt{q^2+TT'}}, \label{eq:solution_saddle}
		\end{equation}
		where $\beta \equiv 1/T$, $\beta' \equiv 1/T'$, and $\Delta \beta \equiv \beta'- \beta$. 
		We note that our perturbation is not valid in the limit $q\rightarrow \pm \infty$ because of the singularity of the scaled cumulant generating function.
		By substituting Eq.~{(\ref{eq:solution_saddle})} into Eq.~{(\ref{eq:large_deviation})}, we obtain 
		\begin{align}
			\lim_{t\rightarrow \infty}\frac{1}{t}\ln{P(q,t)} =& 1+\frac{\Delta \beta q}{2} -\frac{\beta+\beta'}{2}\sqrt{q^2+TT'}\notag\\
															 &+	\sum_{n=2}^\infty \frac{K _{2n}}{n!}\left[\frac{1}{4T ^2}\left( q+\frac{2q^2+T \Delta T}{2\sqrt{q^2+TT'}}\right)\right]^n
															 +	\sum_{n=2}^\infty \frac{K'_{2n}}{n!}\left[\frac{1}{4T'^2}\left(-q+\frac{2q^2-T'\Delta T}{2\sqrt{q^2+TT'}}\right)\right]^n, \label{eq:P}
		\end{align}
		which implies the generalized fluctuation theorem (8).
				
\section{Concluding remarks}
		In this paper, we have studied heat conduction induced by non-Gaussian noises from two athermal environments.
		As a result, we found new terms in the Fourier law and the heat fluctuation theorem, 
		which implies that the heat current can be induced by the non-Gaussianity of athermal fluctuations. 
		We have also discussed that the zeroth law of thermodynamics is not straightforwardly valid for athermal systems. 
		Our numerical results are not consistent with the conventional Fourier law and the fluctuation theorem, but consistent with the analytical results obtained in this paper.

		Our theory is the first departure from the Gaussian stochastic thermodynamics toward a universal theory of nonequilibrium statistical mechanics in the presence of non-Gaussian noises. 
		It is interesting to investigate if the generalized Fourier law and fluctuation theorem obtained in this paper would hold in a much broader class of athermal heat conduction. 

\begin{acknowledgments}
		We are grateful to K. Sekimoto, K. Kawaguchi, T. Nemoto, S. Ito, and H. Takayasu for valuable discussions.
		The numerical calculations were carried out on SR16000 at YITP in Kyoto University. 
		This work was supported by the Grants-in-Aid for Japan Society for Promotion of Science (JSPS) Fellows (Grant No. 24$\cdot$3751), the Grant-in-Aid for Research Activity Start-up (Grant No. 11025807) 
		and the Global COE Program, ``The Next Generation of Physics, Spun from Universality and Emergence" from the Ministry of Education, Culture, Sports, Science and Technology (MEXT) of Japan. 
\end{acknowledgments}
\appendix

\section{Cumulant functional and $n$-points delta functions}
		We discuss the relationship between the cumulant functional of white non-Gaussian noise $\hat \xi(t)$ and the $n$-points delta functions. 
		Let us introduce the characteristic functional $\mathcal{G}[v]$ and the cumulant functional $\mathcal{H}[v]$~\cite{Hanggi} as:
		\begin{equation}
			\mathcal{G}[v] \equiv \langle e^{i\int_0^t ds \hat \xi(s)v(s)}\rangle,\>\>\>
			\mathcal{H}[v] \equiv \ln {\mathcal{G}[v]},
		\end{equation}
		where $v(s)$ is an arbitrary function. 
		The $n$-th order moments $\langle \hat \xi(t_1)\dots \hat \xi(t_n)\rangle$ and the $n$-th order cumulants $\langle \hat \xi(t_1)\dots \hat \xi(t_n)\rangle_c$ can be respectively written as
		\begin{align}
			\langle \hat \xi(t_1)\dots \hat \xi(t_n)\rangle 	&\equiv \frac{\delta^n \mathcal{G}[v]}{\delta iv(t_1)\dots \delta iv(t_n)}\Bigg|_{v=0},\\
			\langle \hat \xi(t_1)\dots \hat \xi(t_n)\rangle_c 	&\equiv \frac{\delta^n \mathcal{H}[v]}{\delta iv(t_1)\dots \delta iv(t_n)}\Bigg|_{v=0}.
		\end{align}
		It is known that there are relations between the moments and the cumulants~\cite{Gardiner}.
		In particular, the forth cumulant can be written as 
		\begin{align}\label{eq:cumulant_moment}
			\langle \hat\xi(t_1)\hat\xi(t_2)\hat\xi(t_3)\hat\xi(t_4) \rangle_c 	=\langle \hat\xi(t_1)\hat\xi(t_2)\hat\xi(t_3)\hat\xi(t_4) \rangle
																				-\langle \hat\xi(t_1)\hat\xi(t_2)\rangle \langle\hat\xi(t_3)\hat\xi(t_4) \rangle
																				-\langle \hat\xi(t_1)\hat\xi(t_3)\rangle \langle\hat\xi(t_2)\hat\xi(t_4) \rangle
																				-\langle \hat\xi(t_1)\hat\xi(t_4)\rangle \langle\hat\xi(t_2)\hat\xi(t_3) \rangle
		\end{align}
		for $\langle\hat \xi(t)\rangle=0$.  
		We note that the fourth cumulant can be decomposed into the second cumulant only in the case of the Gaussian noise. 
		According to the L\'evy-It\^o decomposition~\cite{Gardiner,Ito}, 
		the cumulant functional can be transformed into the standard form of L\'evy processes 
		\begin{equation}
			\mathcal{H}[v] = \int_0^t ds\left[ iav(s) - \frac{\sigma^2v^2(s)}{2} + \int_{-\infty}^{+\infty} dz \left(e^{iv(s)z} - 1\right)w(z)\right], 
		\end{equation}
		where $a$ and $\sigma^2$ are arbitrary constants, $w(z)$ is a transition rate function. 
		
		We next introduce the $n$-points delta functions as~\cite{Kleinert}
		\begin{align}
			\delta_n (t_1,\dots,t_n) &= \begin{cases}
											\infty & (t_1=\dots=t_n)\\
											0 & (\mathrm{otherwise})
										\end{cases},\\
			\int_{-\infty}^{\infty}dt_2\dots dt_n &\delta_n(t,t_2,\dots t_n) = 1 \label{eq:def_n_delta}.
		\end{align}
		We assume the symmetric property for the delta function $\delta_n(t_1,\dots t_n) =\delta_n(s_1,\dots,s_n)$,
		where $\{ s_1,\dots s_n\}$ is an arbitrary permutation of $\{t_1,\dots,t_n\}$. 
		The following equality can be derived from this symmetric property:
		\begin{equation}\label{eq:edge_delta}
			\int_{0}^tdt_2dt_3dt_4 \delta_4(t,t_2,t_3,t_4) = \frac{1}{4}. 
		\end{equation}
		The derivation of Eq.~{(\ref{eq:edge_delta})} is as follows.  
		By definition, we obtain
		\begin{equation}
			\int_{0}^t dt_1dt_2dt_3dt_4 \delta_4(t_1,t_2,t_3,t_n) = t.
		\end{equation}
		By differentiating the left and right hand sides with respect to $t$, we obtain 
		\begin{equation}
			\int_{0}^t dt_2dt_3dt_4 \delta_4(t,t_2,t_3,t_4) + \int_0^t dt_1dt_3dt_4 \delta_4(t_1,t,t_3,t_4)+\dots + \int_0^t dt_1dt_2dt_3 \delta_4(t_1,t_2,t_3,t)= 1. 
		\end{equation}
		The symmetric property of the delta function leads to the following equality
		\begin{equation}
			4\int_{0}^tdt_2dt_3dt_4 \delta_4(t,t_2,t_3,t_4) = 1,
		\end{equation}
		which implies Eq.~(\ref{eq:edge_delta}). 
		Similar equalities can be derived using the parallel techniques. 
		
		We can represent the functional derivatives of the cumulant functional with the $n$-points delta functions:  
		\begin{align}
			\frac{\delta^2 \mathcal{H}[v]}{\delta iv(t_1)\delta iv(t_2)}\Bigg|_{v=0} &= 2T\delta_2(t_1,t_2),\notag\\
			\frac{\delta^n \mathcal{H}[v]}{\delta iv(t_1)\delta iv(t_2)\dots \delta iv(t_n)}\Bigg|_{v=0} &= K_n\delta_n(t_1,t_2,\dots,t_n)\notag
		\end{align} 
		where 
		$K_2 = 2T\equiv \sigma^2 + \int z^2w(z)dz$ and $K_n \equiv \int z^{n} w(z)dz$ for $n \geq 3$. 		
		We note that the assumption of the symmetry of the delta function is consistent with that of the mixed functional derivative 
		\begin{equation}
			\frac{\delta^n \Phi[v]}{\delta iv(t_1)\dots \delta iv(t_n)}\Bigg|_{v=0} = \frac{\delta^n \Phi[v]}{\delta iv(s_1)\dots \delta iv(s_n)}\Bigg|_{v=0},
		\end{equation}
		where $\{ s_1,\dots s_n\}$ is an arbitrary permutation of $\{t_1,\dots,t_n\}$. 

\section{The $\ast$ integral}
		We briefly review the formulation of the $\ast$ integral~\cite{Kanazawa}. 
		The main idea of the $\ast$ integral is to take a white noise limit of a colored noise in order to remove the singularity of the white noise. 
		Let $\hat x(t)$ be an arbitrary stochastic variable. 
		The $\ast$ integral for an arbitrary function $f(\hat x(t))$ is defined as a white noise limit of a colored noises: 
		\begin{equation}
			\int_0^t ds \hat \xi(s)\ast f(\hat x(s)) \equiv 
			\lim_{\varepsilon \rightarrow +0}\lim_{\Delta t\rightarrow +0}\sum_{i=0}^{N-1} \Delta t \hat \xi_\varepsilon(t_i) f(\hat x(t_i)), 
		\end{equation} 
		where $\Delta t\equiv t/N$, $t_i \equiv i\Delta t$, and
		$\hat \xi_\varepsilon(t)$ is a colored noise with a finite correlation time $\varepsilon$.
		An explicit definition of $\hat \xi_\varepsilon(t)$ is given by
		\begin{equation}
			\hat \xi_\varepsilon(t) \equiv \frac{1}{\varepsilon}\int_t^{t+\varepsilon} ds\hat\xi(s).
		\end{equation}
		The $\ast$ integral is a generalization of the Stratonovich integral, and is the same as the Stratonovich integral for Gaussian processes.
		An advantage of the $\ast$ calculus lies in the fact that the chain rule holds even for non-Gaussian processes~\cite{Kanazawa}. 
		The $\ast$ integral is applicable to the definition of heat in stochastic energetics~\cite{Sekimoto1,Sekimoto2,Sekimoto3,Kanazawa}. 

		The $\ast$ integral can be transformed into the It\^o integral, which is a crucial technique for the derivations of Eqs. (6) and (8) in the main text. 
		Let us assume a Langevin equation and the corresponding stochastic heat current respectively as 
		\begin{equation}
			\frac{d\hat z}{dt} = -2\frac{dU(\hat z)}{d\hat z} + \hat \xi +\hat \eta,\>\>\>\>
			\frac{d\hat Q}{dt} = -\left(\frac{dU(\hat z)}{d\hat z}\right)^2 + \frac{dU(\hat z)}{d\hat z}\ast \xi,
		\end{equation}
		where $U(\hat z)$ is a potential functions, and $\hat \xi$ and $\hat \eta$ are white non-Gaussian noises. 
		The $\ast$ integrals for an arbitrary function $f(\hat z,\hat Q)$ can be transformed into the It\^o integrals as 
		\begin{align}
			d\hat L   	\ast f(\hat z,\hat Q) &= \sum_{n=0}^\infty \frac{\left( d\hat L \right)^{n+1}}{(n+1)!}
						\cdot \left[\frac{\partial }{\partial \hat z}+\frac{dU(\hat z)}{d\hat z}\frac{\partial }{\partial \hat Q}\right]^nf(\hat z,\hat Q) \label{eq:transform1}\\
			d\hat L'	\ast f(\hat z,\hat Q) &= \sum_{n=0}^\infty \frac{\left( d\hat L'\right)^{n+1}}{(n+1)!}
						\cdot \left[\frac{\partial }{\partial \hat z}\right]^nf(\hat z, \hat Q)\label{eq:transform2}
		\end{align}
		where the symbol $\cdot$ denotes the It\^o integral, and the L\'evy processes $\hat L(t)$, $\hat L'(t)$ are respectively defined by $\hat L(t)\equiv \int_0^t ds\hat \xi(t)$ and $\hat L'(t)\equiv \int_0^t ds\hat \eta(t)$. 
		These equations can be derived as follows. 
		According to Ref.~\cite{Hanggi}, 
		\begin{align}
			&\langle \hat \xi \ast f(\hat z,\hat Q)\rangle 
			= \lim_{t\rightarrow \infty}\langle \hat \xi_{\epsilon}f(\hat z,\hat Q)\rangle\notag\\
			&= \sum_{n=0}^\infty\frac{K_{n+1}}{(n+1)!} \left<\frac{\delta^n f(\hat z,\hat Q)}{\delta \hat \xi^n}\right>\notag\\
			&= \sum_{n=0}^\infty\frac{\langle d\hat L^{n+1}/dt \rangle}{(n+1)!}
			\left<\left[\frac{\partial }{\partial \hat z} + \frac{dU(\hat z)}{d\hat z}\frac{\partial }{\partial \hat Q}\right]^n f(\hat z,\hat Q)\right>,
		\end{align}
		where $\langle d\hat L^n\rangle = K_ndt$, $\delta \hat z/ \delta\hat \xi = 1$, and $\delta \hat Q/\delta \hat \xi = dU/d\hat z$. This equation can be rewritten as
		\begin{equation}
			\langle d\hat L \ast f(\hat z,\hat Q)\rangle = \sum_{n=0}^\infty\left<\frac{d\hat L^{n+1}}{(n+1)!}\cdot 
			\left[\frac{\partial }{\partial \hat z} + \frac{dU(\hat z)}{d\hat z}\frac{\partial }{\partial \hat Q}\right]^n f(\hat z,\hat Q)\right>.
		\end{equation}
		Because this equality holds for an arbitrary function $f(\hat z,\hat Q)$, we obtain Eq.~{(\ref{eq:transform1})}. In a parallel calculation, we obtain Eq.~{(\ref{eq:transform2})}.

\section{Weakly quartic potential}
		In this appendix, we discuss a correction term to the Fourier law for a weakly quartic potential with non-Gaussian noises. 
		Let us consider a system with a weakly quartic potential $U(\hat z)=\hat z^2/2 + \epsilon\hat z^4/4$, where $\epsilon$ is a small constant. 
		Here we do not assume that the temperature difference $\Delta T$ and the non-Gaussian properties $\{ K_n\}_{n\geq 3}$ are also small. 
		In the first order perturbation in terms of $\epsilon$, we obtain a correction term to the Fourier law as
		\begin{equation}\label{eq:appendix_GFL}
			J = -\kappa\Delta T - \kappa'\Delta K_4+O(\epsilon^2), 
		\end{equation} 
		where 
		$\kappa  \equiv (1/2)\left[ 1+\left\{3\epsilon(T+T')/2\right\}\right]$ and $\kappa' \equiv \epsilon/8$.
		We note that only the fourth cumulant difference $\Delta K_4$ appears in the rhs of Eq.~{(\ref{eq:appendix_GFL})} as the correction term because 
		$d^nU/d\hat z^n = 0$ with $n=3$ and $n\geq 5$. 
		This result is consistent with Eq.~{(\ref{eq:heat_flux_general})} when $\Delta T$ and $\{ K_n\}_{n\geq 3}$ are small. 
		A similar result to Eq.~{(\ref{eq:appendix_GFL})} was obtained for an underdamped system with a weakly quartic potential~\cite{Morgado2}. 
		We note that the zeroth law of thermodynamcis is not straightforwardly valid because the condition of $J=0$ in Eq.~{(\ref{eq:appendix_GFL})} explicitly depends on the properties of the heat conductor. 
		However, we can introduce the device-dependent indicator $\mu_\epsilon(T,K_4) \equiv T/2 +3\epsilon T^2/4 + \epsilon K_4/8$ to show the transitive relation if we fix the contact device,
		where $\mu_\epsilon$ characterize the direction of heat current as $J=\mu_\epsilon(T,K_4)-\mu_\epsilon(T',K'_4)$.
		
		Equation~{(\ref{eq:appendix_GFL})} can be derived as follows. 
		We assume that the solution of Eq.~{(\ref{eq:single_eq})} is expanded as $\hat z(t) = \hat z_0(t) + \epsilon\hat z_1(t) + O(\epsilon^2)$, 
		where $\hat z_0(t)$ and $\hat z_1(t)$ respectively satisfy 
		\begin{equation}\label{eq:appendix_z}
			\frac{d\hat z_0}{dt} + 2\hat z_0 = \hat \xi + \hat \eta,\>\>\>
			\frac{d\hat z_1}{dt} + 2\hat z_1 = -2\hat z^3_0.
		\end{equation}
		By solving Eq.~{(\ref{eq:appendix_z})}, we obtain the explicit solution 
		\begin{equation}\label{eq:appendix_solution}
			\hat z(t) 	= 	\int_0^tds_1e^{-2(t-s_1)}\left(\hat \xi_1+\hat \eta_1\right)-2\epsilon\int_0^t ds_1 e^{-2(t-s_1)}\int_0^{s_1} \prod _{i=2}^4 ds_i e^{-2(s_1-s_i)}\left(\hat \xi_i + \hat \eta_i\right), 
		\end{equation}
		where we denote $\hat \xi_n$ and $\hat \eta_n$ by $\hat \xi(s_n)$ and $\hat \eta(s_n)$ with a positive integer $n$, respectively. 
		From straightforward calculations, we obtain 
		\begin{align}
			&\langle \hat z\ast \hat \xi\rangle_{\rm{SS}} = T + O(\epsilon^2),\label{eq:ghf_3} \\
			&\langle \hat z^3 \ast \hat \xi\rangle_{\rm{SS}} = \frac{3T(T+T')}{2} + \frac{K_4}{4}+O(\epsilon),\label{eq:ghf_4} \\
			&\langle z^2\rangle_{\rm{SS}} = \frac{T+T'}{2}-\frac{3\epsilon(T+T')^2}{4}-\frac{\epsilon(K_4+K'_4)}{8}+O(\epsilon^2),\label{eq:ghf_1}\\
			&\langle \hat z^4\rangle_{\rm{SS}} = \frac{3(T+T')^2}{4} + \frac{K_4+K'_4}{8} + O(\epsilon).\label{eq:ghf_2}
		\end{align}
		From Eqs. {(\ref{eq:ghf_3})} - {(\ref{eq:ghf_2})} and {(\ref{eq:stochastic_heat})}, we then obtain 
		\begin{align}
			J 	&= \langle \hat z \ast \hat \xi \rangle_{\rm{SS}} + \epsilon\langle \hat z^3\ast \xi\rangle_{\rm{SS}}-\langle \hat z^2\rangle_{\rm{SS}} - 2\epsilon\langle \hat z^4\rangle_{\rm{SS}} +O(\epsilon^2)\notag\\
				&= -\frac{1}{2}\left[1+\frac{3\epsilon(T+T')}{2}\right]\Delta T - \frac{\epsilon}{8}\Delta K_4 +O(\epsilon^2),\label{eq:ghf_5}
		\end{align}
		which implies Eq.~{(\ref{eq:appendix_GFL})}.
		
		Here we show the explicit derivations of Eqs.~{(\ref{eq:ghf_3})} - {(\ref{eq:ghf_2})}. 
		$\langle \hat z(t)\ast \hat \xi(t)\rangle $ for an arbitrary $t$ can be written as
		\begin{align}
			\langle \hat z(t)\ast \hat \xi(t)\rangle&= \int_0^t dse^{-2(t-s)}\langle\hat \xi(t)\hat \xi(s)\rangle - 2\epsilon \left<\int_0^t dse^{-2(t-s)}\hat \xi(t) \int_0^s \prod_{i=1}^3ds_i e^{-2(s-s_i)} 
														(\hat \xi_i+\hat \eta_i)\right> + O(\epsilon^2)\notag\\
													&= T +O(\epsilon^2),
		\end{align}
		which implies Eq.~{(\ref{eq:ghf_3})}. 
		$\langle \hat z^3(t)\ast \hat \xi(t)\rangle$ for an arbitrary $t$ can be written as
		\begin{align}
			\langle \hat z^3(t)\ast \hat \xi(t)\rangle  =&	\left<\hat \xi(t)\int_0^t\prod_{i=1}^3 ds_i e^{-2(t-s_i)}(\hat \xi_i+\hat \eta_i)\right>+O(\epsilon)\notag\\
														=&	6T(T+T')\int_0^tdse^{-4(t-s)} + \frac{K_4}{4}+O(\epsilon) \notag\\
														=&	\frac{3T(T+T')}{2}\left[1-e^{-4t}\right]+\frac{K_4}{4} +O(\epsilon),
		\end{align}
		which implies Eq.~{(\ref{eq:ghf_4})}.
		$\langle\hat z^2(t)\rangle$ and $\langle \hat z^4(t) \rangle$ are explicitly given by
		\begin{widetext}
			\begin{align}
				\langle\hat z^2(t)\rangle	\!=&\! 	\int_0^t \!\!\left<\prod_{i=1}^2\!ds_i e^{-2(t-s_i)}\!\!\left(\hat \xi_i\!+\!\hat \eta_i\right)\!\!\right> \!
											-	\!\left<\!\!4\epsilon\!\! \int_0^t \!\!ds_1\int_0^t\!\! ds_2 e^{-2(t-s_1)-2(t-s_2)}\!\!\left(\hat \xi_1\!+\!\hat \eta_1\!\right)\!\!\int_0^{s_2} \!\!\prod_{i=3}^5 \!ds_i e^{-2(s_2-s_i)}
												\left(\hat \xi_i\!+\!\hat \eta_i\right) \!\!\right>
											+ O(\epsilon^2).\notag\\
											=&	2(T+T')\int_0^t ds e^{-4(t-s)} -48\epsilon(T+T')^2\int_0^t ds_2\int_0^{s_2}ds_1\int_0^{s_2} ds_3 e^{-2(t-s_1)-2(t-s_2)-2(s_2-s_1)-4(s_2-s_3)}\notag\\
											&	-4\epsilon(K_4+K'_4)\int_0^t ds_2\int_0^{s_2}ds_1e^{-2(t-s_1)-2(t-s_2)-6(s_2-s_1)} + O(\epsilon^2)\notag\\
											=& \frac{T+T'}{2}\left[1-e^{-4t}\right]-\frac{3\epsilon (T+T')^2}{4}\left[1-8te^{-4t}-e^{-8t}\right]-\frac{\epsilon(K_4+K'_4)}{8}\left[1-e^{-4t}\right]^2 + O(\epsilon^2),\label{eq:z_square}\\
				\langle\hat z^4(t)\rangle 
						=& \int_0^t\left< \prod_{i=1}^4 ds_i e^{-2(t-s_i)}\left(\hat \xi_i+\hat \eta_i\right)\right>+O(\epsilon)\notag\\
				=& 12(T+T')^2\left[\int_0^t ds e^{-4(t-s)}\right]^2
						+(K_4+K'_4)\int_0^tdse^{-8(t-s)}+O(\epsilon)\notag\\
				=&\frac{3(T+T')^2}{4}\left[1-e^{-4t}\right]^2 + \frac{K_4+K'_4}{8}[1-e^{-8t}]+O(\epsilon), \label{eq:z_fourth}
			\end{align}
		\end{widetext}
		where we have used Eqs.~{(\ref{eq:appendix_solution})} and~{(\ref{eq:edge_delta})}. 
		Eqs.~{(\ref{eq:z_square})} and (\ref{eq:z_fourth}) respectively imply Eq.~{(\ref{eq:ghf_1})} and~{(\ref{eq:ghf_2})} in the steady limit $t\rightarrow \infty$.
		
\section{Non-lienar part of the generalized heat fluctuation theorem}
		\begin{figure}
			\centering
			\includegraphics[width=80mm,clip]{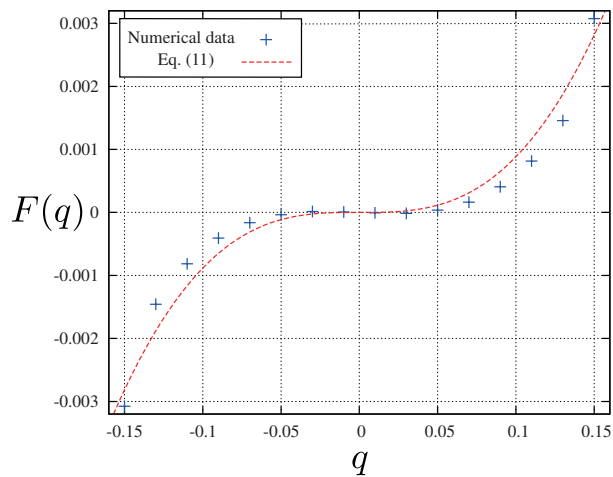}
			\caption{
						Numerical observation of the non-linear effect in the fluctuating function $F(q)$. 
						The cross points indicate the numerical data of $F(q)$ for $t=100$, 
						and the solid line is the theoretical line obtained from Eq.~{(\ref{eq:Modified_FT})}. 
						We perform the Monte Carlo simulation to make the histogram of the heat distribution function, 
						and numerically obtain $\tilde{F}(q,t) \equiv (1/t)\ln{P(q,t)/P(-q,t)}$ for $t=50$ and $t=100$. 
						According to the Richardson extrapolation~\cite{Numerical_recipes}, we have plotted $2\tilde{F}(q,t=100)-\tilde{F}(q,t=50)$ as the fluctuating function $F(q)$ for $t=100$.
						The bin-width for the heat histogram is $0.02$, 
						the time step is $0.0002$, and 
						the number of samples is $7.6 \times 10^8$. 
					}
		\end{figure}
		We have numerically observed the non-linear effect in Eq.~{(\ref{eq:g_FT_P})} in terms of $q$. 
		Figure 4 shows the numerical data of the fluctuating function $F(q)$ for $t=100$ with $T=T'=0.30$, $\lambda=20.0$, and $\lambda'=\infty$.
		Due to large cost of the numerical simulation, we could not observe the convergence of $F(q)$ to our theoretical line (\ref{eq:g_FT_P}) in the limit of $t\rightarrow\infty$.

\end{document}